\documentclass[pre,superscriptaddress,twocolumn,a4paper,showpacs]{revtex4-1}
\usepackage{amssymb}
\usepackage{wasysym}
\usepackage{graphicx}
\usepackage{epsfig}
\usepackage[caption=false]{subfig}
\usepackage{float}
\usepackage[usenames, dvipsnames]{color}
\begin{document}

\title{Impact of  centrality on cooperative processes}

\author{Sandro M. Reia}
\affiliation{Instituto de F\'{\i}sica de S\~ao Carlos,
  Universidade de S\~ao Paulo,
  Caixa Postal 369, 13560-970 S\~ao Carlos, S\~ao Paulo, Brazil}
  
  \author{Sebastian Herrmann}
\affiliation{Department of Information Systems and Business Administration, Johannes Gutenberg-Universit\"at, 
Jakob Welder-Weg 9, 55128 Mainz, Germany}
             
\author{Jos\'e F.  Fontanari}
\affiliation{Instituto de F\'{\i}sica de S\~ao Carlos,
  Universidade de S\~ao Paulo,
  Caixa Postal 369, 13560-970 S\~ao Carlos, S\~ao Paulo, Brazil}

\begin{abstract}
 The solution of  today's complex problems requires the grouping of task forces whose members are   usually connected remotely over
 long physical distances and different time zones. Hence, understanding the effects of imposed communication patterns (i.e., who can communicate with whom) on group performance is important.  Here, we  use an agent-based model to explore the influence of the betweenness centrality of the nodes on the time the group requires to find the global maxima of NK-fitness landscapes.   
The agents cooperate by broadcasting  messages,  informing on their fitness to their  neighbors, and use
this information  to copy the more successful agents in their neighborhood. We find that for easy tasks (smooth landscapes), the topology of the communication network has no effect on the  performance of the group, and that the more central nodes are the most likely to find the global maximum first.  For difficult tasks (rugged landscapes), however, we find a positive correlation between the  variance of the betweenness  among the network nodes and the group performance. For these tasks, the performances of individual nodes are strongly influenced by the agents dispositions to cooperate  and by the particular realizations of the rugged landscapes.
\end{abstract}

\maketitle

\section{Introduction}\label{sec:intro}

Problem solving by task groups  represents a substantial portion  of the economy  of developed countries nowadays \cite{Page_07}.  Among
the  work relationship  issues that emerge in this situation, the most important is perhaps that  of intra-group communication. In fact, the question ``What effect do communication patterns have upon the operation of groups?'' prompted a series of experimental studies in the 1950s, which produced somewhat conflicting conclusions \cite{Bavelas_50,Leavitt_51,Heise_51,Guetzkow_55,Shaw_54,Mulder_60}.
Of particular interest  is the case of imposed communication patterns, which happens in the military and industrial organizations, for instance,  and  in which the researchers determine  who can communicate with whom, thus excluding a priori  the alternative of self-organization of the group members.

Here we address the issue of the influence of a fixed communication pattern on  group performance, as measured by the time the group needs to find the solution of a task. Already in  the pioneer studies of the 1950s, the concept of centrality has emerged as the chief (but not the sole) determinant of the differences in  performance of the various group organizations  \cite{Bavelas_50,Leavitt_51}. Centrality or, more  precisely, betweenness centrality is a concept of the importance of a member for the diffusion process in a network along the shortest paths. Hence, betweenness is a measure of the availability of the information necessary for solving the task \cite{Freeman_77}. In fact, a typical finding of those studies was  that  the most central position in a pattern (e.g., in a wheel), which is located on many shortest-path information flows between all other positions, is most likely to hit the solution first \cite{Leavitt_51}. 

Rather than studying small groups of human subjects as in those seminal works, we consider agent-based simulations, aiming at offering a more complete understanding of the interplay between the  centrality of the communication patterns and the  complexity of the task. Even though it is debatable that conclusions drawn from such an approach may apply to groups of human workers (see, e.g.,  \cite{Mason_12}), they certainly hold for distributed computational systems that are ubiquitous in today's society \cite{Huberman_90,Clearwater_91,Clearwater_92}.

In particular, we consider a distributed cooperative problem solving model
 in which agents cooperate by broadcasting messages, informing on their partial success towards the completion of the goal. The agents use this information to copy parts of the tentative answer exhibited by the more successful agents  in their influence networks \cite{Fontanari_14}. 
Since copying is an essential ingredient of social learning (i.e., learning through observation), and is central to the remarkable success of our species \cite{Blackmore_00,Rendell_10}, we expect that our conclusions may be of relevance to the organization of real-world task-groups.
The    parameters of the model  are the number of agents  in the system  $L$ and the copy propensity  $p \in \left [0,1 \right ]$ that is the same for all agents. 

The relevant network metric to our study is  the betweenness centrality, which  measures a node's centrality in a communication pattern \cite{Freeman_77}. 
Although there are many other measures of centrality, such as random walk betweenness centrality \cite{Newman_05}, eigenvector centrality \cite{Seeley_49}  and knotty-centrality \cite{Shanahan_12}, to mention only a few, here we focus on the  betweenness centrality, which implicitly assumes that information flows between nodes  through the shortest paths. Thus, our  findings can be compared
with the results from the literature which used this centrality measure \cite{Bavelas_50,Leavitt_51,Mason_12}.
In order to single out  the influence of the betweenness centrality on the group performance, we follow \cite{Mason_12} and  fix the group size to $L=16$  and the degrees of the nodes  to $k=3$ (see Fig.\ \ref{fig:1}). In the rest of the paper we will use the terms communication pattern and network interchangeably. 
The task posed to the agents is to find the unique global maximum of a fitness landscape, whose state space is much larger than the group size $L$. The  difficulty  of  the task is gauged by the number and distribution  of local maxima in the landscape.

We find that for easy tasks (i.e., for  landscapes  without local maxima) the network topology has no apparent effect on the group performance but for difficult tasks,  the strength of the  performance is  associated to the variance of the betweenness centrality among the network nodes. The network which maximizes this variance exhibits a hierarchical organization with a central node and a modular structure (network A in Fig.\ \ref{fig:1}).  It  is  interesting that such an organization performs better than a more equalitarian pattern, in which the betweenness centrality of all nodes is maximized (network B in Fig.\ \ref{fig:1}). Moreover, we find that the best performances are achieved by the so-called  inefficient networks, which are characterized by long average path lengths that delay the propagation of information  through the network. 
This is because in rugged landscapes, the information on fitness exchanged by the agents  is often misleading, hinting to the locations  of  local maxima, rather than to the position of the global maximum of the fitness landscape.

In addition, a more detailed consideration of the performance of the nodes shows that for easy tasks, the chance that a node finds the answer first is positively correlated with its betweenness centrality. For difficult tasks, however, the chance of a node hitting the solution depends on the copy propensity $p$ of the agents: for small $p$, all agents are roughly equiprobable of finding the solution. Near the value of $p$ that optimizes the group performance, the central agents perform better. For large $p$,  the more peripheral nodes  have a better chance to  get the answer first.

The rest of this paper is organized as follows. In Section \ref{sec:NK}, we offer  an outline of the NK model of   rugged fitness landscapes  \cite{Kauffman_87}, which we use to represent the tasks presented to the agents.
 The  behavioral rules that guide the agents in their searches for the global maximum of the  landscapes are explained in Section \ref{sec:model}.  The four fixed  communication patterns the agents use to exchange information on their tentative solutions  are introduced in Section \ref{sec:nets}. 
In  Section \ref{sec:res}, we present and analyze the results of our simulations, emphasizing the comparative performance  between the different patterns.  Finally, Section \ref{sec:disc} is reserved to our concluding remarks.


\section{Task}\label{sec:NK}

The task posed to a system of $L$  agents $i=1,\ldots,L$ is  to find the unique global maximum of  a fitness landscape  using   the
NK model  \cite{Kauffman_87}.  The NK model is the paradigm  for problem representation in organizational theory  \cite{Levinthal_97,Rivkin_00,Lazer_07,Herrmann_14,Billinger_13}, since it allows the  tuning  of the ruggedness of the landscape -- and hence of the difficulty of the task -- by changing the integer parameters  $N$ and $K$. More pointedly,  an NK landscape is defined in the space of binary strings of length $N$, and so this parameter determines the size of the state space, namely, $2^N$.  The other parameter  $K =0, \ldots, N-1$ is the degree of epistasis that influences the ruggedness of  the landscape for fixed $N$.  
For $K=0$, the (smooth) landscape has a single maximum, whereas for $K=N-1$, the (uncorrelated) landscape  has on the average  $2^N/\left ( N + 1 \right)$ maxima with respect to single bit flips, and the NK model reduces to the Random Energy model \cite{Derrida_81,Saakian_09}.
Finding   the global maximum of the NK model for $K>0$ is an NP-complete problem \cite{Solow_00}, which  means that the time required to solve {\it all}  realizations of that landscape using any currently known deterministic algorithm increases exponentially fast with the length $N$ of the strings \cite{Garey_79}. We refer the reader to the original paper by Kauffman and Levin for details on the procedure to generate a random realization of an NK landscape \cite{Kauffman_87}.

\begin{figure}[!ht]
\centering
   \subfloat{\includegraphics[width=0.23\textwidth]{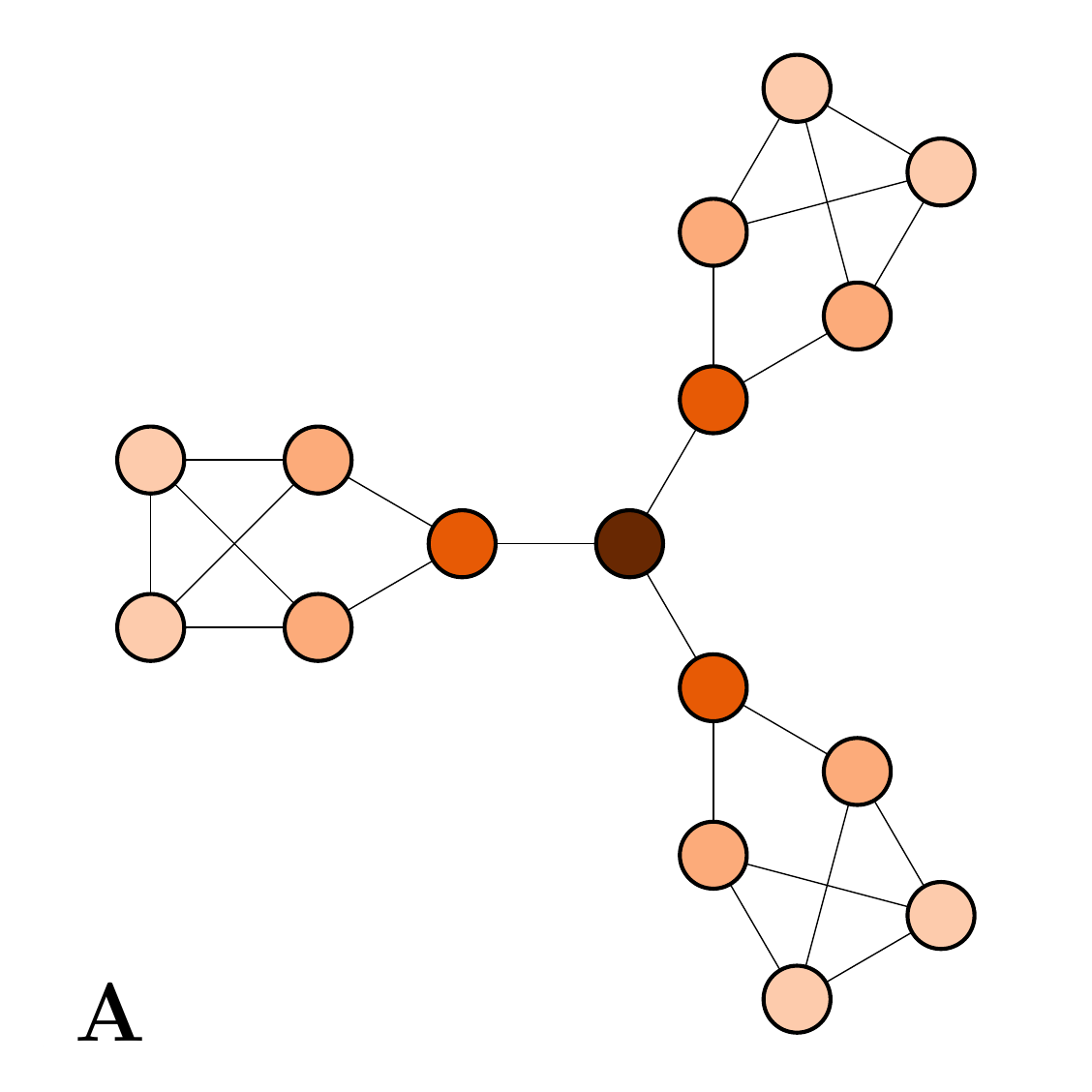}}
   \subfloat{\includegraphics[width=0.23\textwidth]{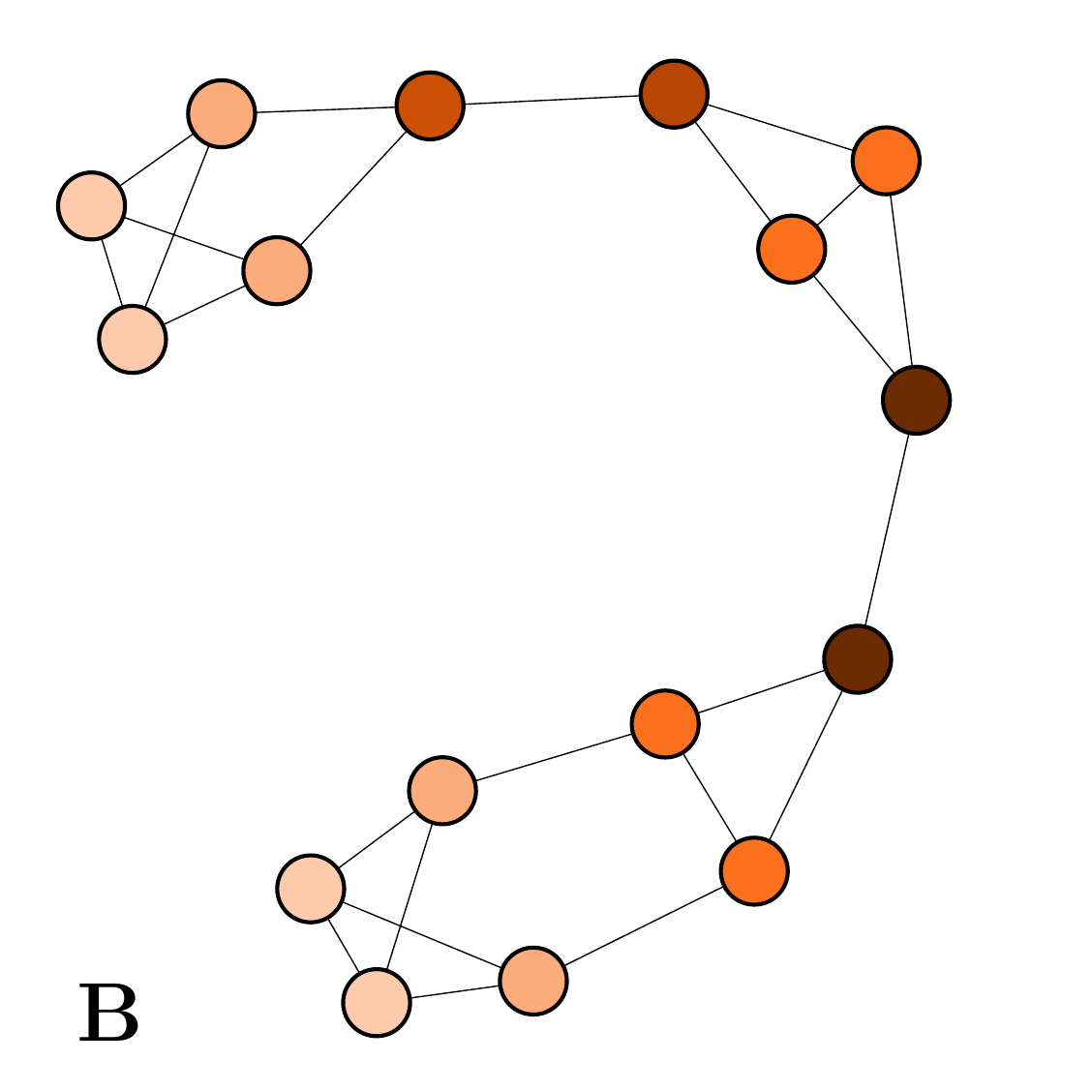}} \\
    	  \subfloat{\includegraphics[width=0.23\textwidth]{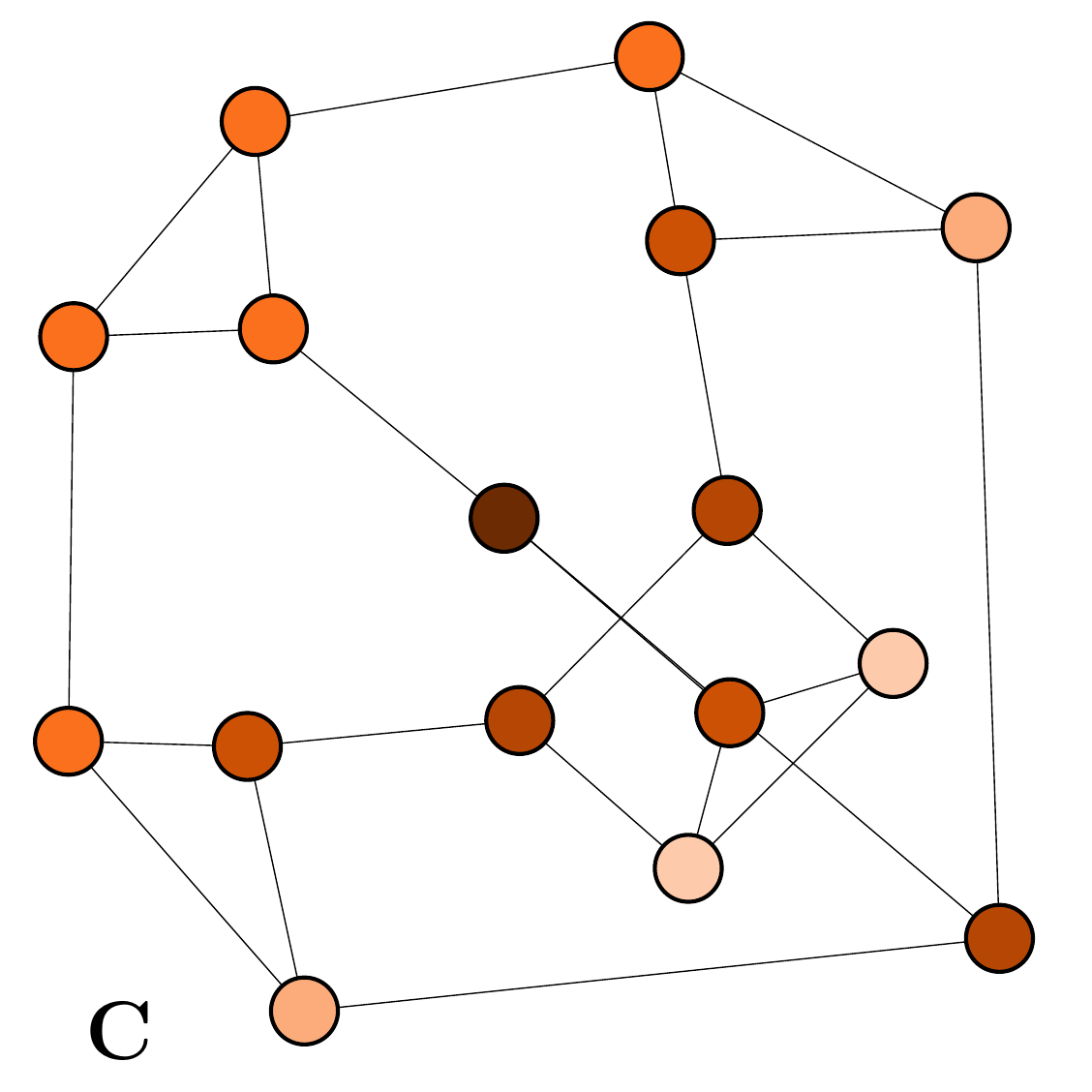}}
    	  \subfloat{\includegraphics[width=0.23\textwidth]{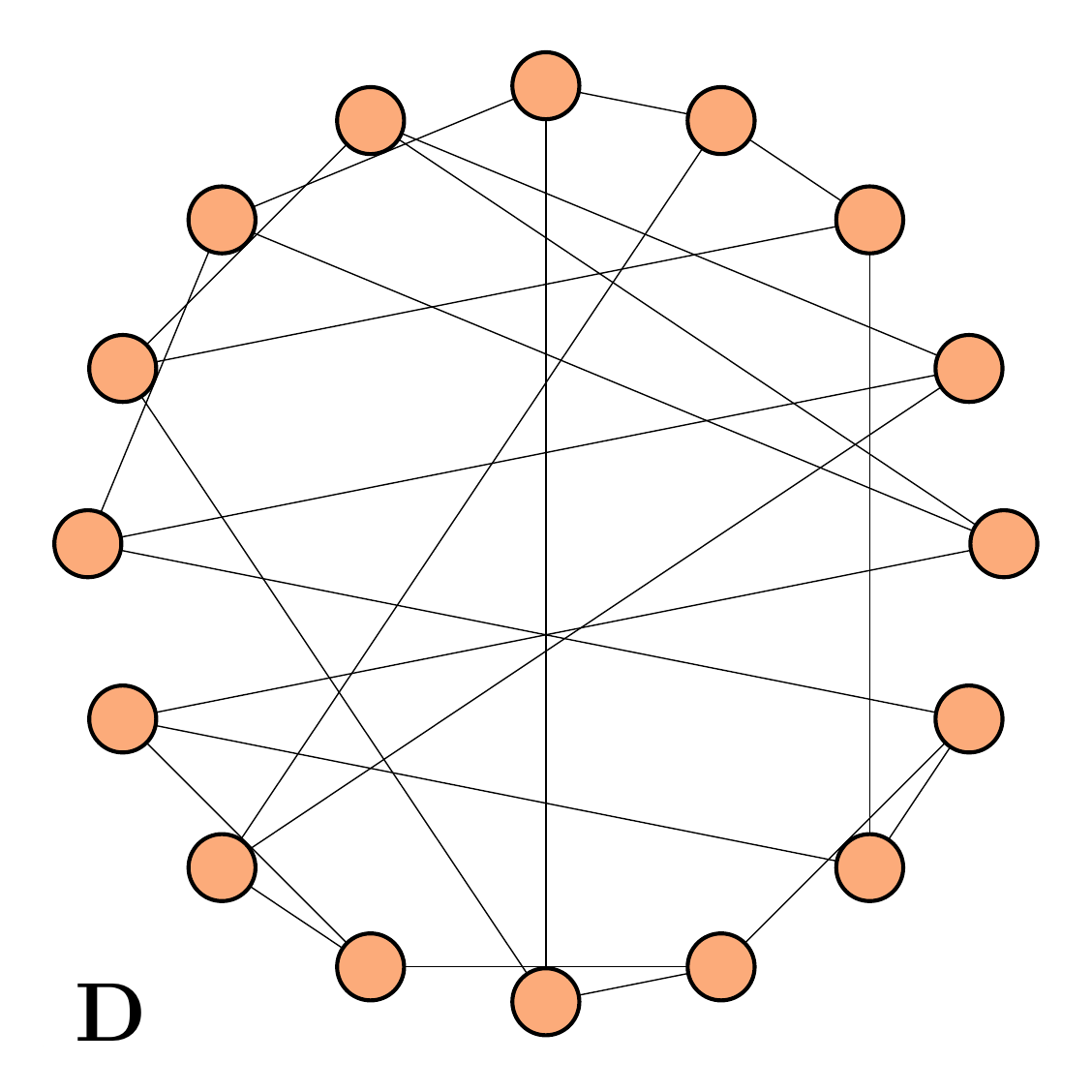}} 
\caption{(Color online)  The four network topologies with $L=16$ nodes and fixed degree $k=3$ used in the computational
experiments.  The darker the shade of a node, the higher its betweenness. Network A is the topology that maximizes both  the maximum betweenness  and the betweenness variance, network B maximizes the average betweenness,  network C is a typical random network  regarding the average  betweenness, and network D  minimizes both the maximum betweenness and the betweenness variance (see Table \ref{table:2}). 
 }
\label{fig:1}
\end{figure}

Since our goal is to compare the performances of cooperative problem-solving systems  using  the four communication patterns shown in Fig.\ \ref{fig:1}, we must guarantee that they search the same realizations of the fitness landscapes, as distinct landscape realizations may differ greatly  in the number  of local maxima for $K>0$.  Thus for each set of the parameters $N$ and $K$, we generate and store 100 landscape realizations,  which we use to test the four patterns. In particular,  we fix the string length  to $N=16$  and allow the degree of epistasis to take on the values $K=0$, $3$, and $7$. 
Table  \ref{table:1} shows the mean number of maxima for each sample of 100 landscapes,  as well as two extreme values, namely,  the minimum and  the maximum number of maxima in the sample.

\begin{table}
\caption{Statistics of the number of maxima in the sample of 100 NK-fitness landscapes with $N=16$ used in the computational experiments.}
\begin{tabular}{c c c c}
\hline
K  & \hspace{1cm} mean  \hspace{1cm} &  min  & max \\ [0.5ex]
\hline
0&  1   &  1 & 1  \\
3 & 84.24  & 43 & 132  \\
 7 & 664.51  & 573 & 770 \\  [1ex]
\hline
\end{tabular}
\label{table:1}
\end{table}
%

\section{Cooperative search}\label{sec:model}

Once the task   is specified, we can set up the representation of the  agents as well as  the rules for their motion on the state space. Clearly, a  convenient model for searching NK landscapes is to represent the agents by binary strings, and so henceforth  we will use the terms agent and string interchangeably.
Initially, the $L$ binary strings are drawn at  random with equal probability for the bits $0$ and $1$. The search begins with the selection of a string (or agent) at random, say string $i$, at time $t=0$. This string  can  move on the state space  through two distinct processes, as described next \cite{Fontanari_14,Fontanari_15}.

The first process, which  happens with probability $p$, is the  imitation  of the model string, which is defined as  the string that exhibits the largest fitness  among the (fixed) subgroup of  strings that  can influence (i.e., are connected to)  string $i$. The  model string and the string $i$   are compared,  and the different bits are singled out.  Then  one of the distinct bits  is selected at random  and flipped,  so that this bit is now the same in both strings.  In the case that string $i$ is identical to the model string,   a randomly chosen bit  of string  $i$  is flipped  with probability one. The second process, which happens with probability $1-p$,  is the elementary move on the state space that consists of picking a  bit  at random  of  string $i$ and flipping it. This elementary move allows the strings to incrementally explore the  entire $2^N$-dimensional state space.    

After string $i$ is updated, we increment the time $t$ by the quantity $\Delta t = 1/L$.  Then another string is selected at random, and the procedure described above is repeated. Note that during the increment from $t$ to $t+1$, exactly  $L$  updates are performed, though not necessarily on $L$ distinct strings.

The  search ends when one of the agents finds the global maximum, and we denote by $t^*$ the halting time. The efficiency of the  search is measured by the total number of string operations necessary  to find that maximum, i.e., $Lt^*$  \cite{Leavitt_51} (see also \cite{Clearwater_91,Clearwater_92})
and so the  computational cost of a search is  defined as $C \equiv L t^*/2^N$, where for convenience we have rescaled $t^*$ by the size of the solution space $2^N$.

The parameter $p \in \left [0,1 \right ]$ is the copy propensity of the agents. The case $p=0$ corresponds to the baseline situation in which    the agents   explore the state space independently of each other.  
The copy or imitation   procedure described above was based on the incremental assimilation mechanism used to study the influence of  external media  \cite{Shibanai _01,Peres_11} in  Axelrod's model  of social influence  \cite{Axelrod_97}. Here we assume that the $L$ agents are identical with respect to their copy propensities (see \cite{Fontanari_16} for the relaxation of this assumption).

The role of imitation on human interactions  was extensively studied by Bandura in the 1960s \cite{Bandura_77}, who concluded that  most human behavior is learned observationally through modeling, i.e., by  observing others and repeating their actions in later similar situations. Most interestingly, Bandura found that the probability of imitation is affected by the characteristics of the observed individual: the higher its perceived status, the more likely it is to be imitated \cite{Bandura_77}. Since the behavioral rules of our agents  concur  with Bandura's findings, their use to  describe the  behavior of human subjects  in
similar collaborative  problem-solving scenarios is justifiable. In addition, we note that similar imitation models have been used to model the strategy of organizations in competitive market situations \cite{Levinthal_97,Rivkin_00}.

A word is in order about a similar agent-based model used in organizational theory  \cite{Lazer_07,Herrmann_14}.   In that model, the agents always copy the fittest string in their neighborhood (i.e., $p=1$, but see below). This move is called exploitation. If the agent is fitter than its neighbors, a single bit is flipped randomly but, differently from our elementary move,  this  change is enacted only if it increases the fitness of the agent. This move is called exploration. In addition, in the case the agent imitates a more successful neighbor, it copies the entire string by changing  many bits  simultaneously. This is then a non-incremental move on the state space. As a result, the search may  permanently get stuck in a  local maximum of the landscape \cite{Lazer_07,Herrmann_14}. This the reason the performance measure in those studies is taken as the average fitness of the group after a fixed search time or as the fraction of searches that  found the global maximum for
unlimited search times.

\section{Communication patterns}\label{sec:nets}

Here we focus on a commonly studied network metric, namely, the betweenness centrality  
 which is a measure of a node's centrality in a communication pattern or network \cite{Freeman_77}. More pointedly, the betweenness centrality of node $i$, which we denote by
 $\mathcal{B}_i$, is given by  the ratio between the number of shortest paths from all nodes to all others that pass through  node $i$ and the number of shortest paths from all nodes to all others regardless of whether they pass through node $i$ or not. Clearly, a node with high betweenness centrality has a large influence on the transfer of information through the network. Hence it is relevant to understand the role of this metric on the performance of distributed cooperative problem-solving systems. We can introduce global metrics as well, such as the average  betweenness centrality $\mu_{\mathcal{B}} = \sum_{i=1}^L \mathcal{B}_i/L$ where $L$ is the number of nodes of the network. The  variance $\sigma_\mathcal{B}^2$ of the betweenness centrality is similarly defined.  For simplicity, henceforth we will refer to the  betweenness centrality as  simply the betweenness, since the other type of betweenness, namely, edge  betweenness \cite{Girvan_02} will not be considered here.
 
 For networks with $L=16$ nodes, each with $k = 3$ neighbors, we can obtain, through an exhaustive search on the space of networks, three networks with special properties regarding the betweenness metric, which are exhibited in Fig.\   \ref{fig:1}. For instance, network A  is the network that maximizes the maximum betweenness among the $L$ nodes. It happens also that this network exhibits the maximum variance (or standard deviation)  of the betweenness  among the nodes.  Network B exhibits the maximum average betweenness. Network D minimizes the maximum betweenness among the $L$ nodes and exhibits also the minimum variance of the betweenness among the nodes.  Actually, all nodes have the same betweenness in this network. Network C is a typical random network, regarding the average betweenness,  with $L=16$ and $k=3$.  To obtain this network we generated a sample of $10^5$ random networks, calculated  the average betweenness of each network, and then the average of the sample: network C was the network whose average betweenness was closest to the sample average. Networks A and B were considered in the study of Ref. \cite{Mason_12}.  Table \ref{table:2} exhibits the 
 average betweenness  ($\mu_\mathcal{B}$) and the standard deviation ($\sigma_\mathcal{B} $) of these four networks. The networks were ordered from high to low values of their betweenness variances.

Another metric of interest used to characterize the communication patterns  is the average path length $\bar{l}$ defined as the average number of steps along the shortest paths for all possible pairs of network nodes. Because it is a measure of the efficiency of the flow of  information on a network \cite{Albert_02}, it  has been used to classify  the  communication patterns as efficient (short path lengths) and
inefficient (long path lengths) \cite{Mason_12}. In that sense, the two networks shown in the upper row of Fig.\ \ref{fig:1} are classified as inefficient  networks and those shown in the lower row as efficient networks (see Table \ref{table:2}). We note that for networks with $L=16$ nodes and fixed degree $k=3$,   networks B and  D have  the largest and the smallest possible  average path lengths, respectively.

\begin{table}
\caption{Summary statistics of the networks' betweenness and average path length for the four topologies used in the computational experiments.}
\begin{tabular}{c c c  c}
\hline
Network  & \hspace{1cm}$ \mu_\mathcal{B} $ \hspace{1cm} &  $~~~~~~~~~\sigma_\mathcal{B} $ \hspace{1cm} & $ \bar{l} $ \\ [0.5ex]
\hline
A &  0.1678   &  0.2118 & 3.35 \\
B & 0.2047 & 0.1862 & 3.87 \\
C &  0.1036 & 0.0348 & 2.45  \\
D & 0.0857  & 0 & 2.2  \\ [1ex]
\hline
\end{tabular}
\label{table:2}
\end{table}
%

\section{Results}\label{sec:res}

For a given communication pattern and for fixed values of the NK model parameters we proceed as follows.
For each realization of a fitness landscape we carry out $10^4$ to $10^5$ searches starting from different initial conditions
(initial strings),  and the resulting average computational cost is then averaged again over $100$  distinct landscapes.  We recall that
the  four networks of Fig.  \ref{fig:1} are tested on the same landscapes. The error bars are smaller than the symbol sizes  in all figures shown
in this section. 

\begin{figure}
\centering
  	\subfloat {\includegraphics[width=0.48\textwidth]{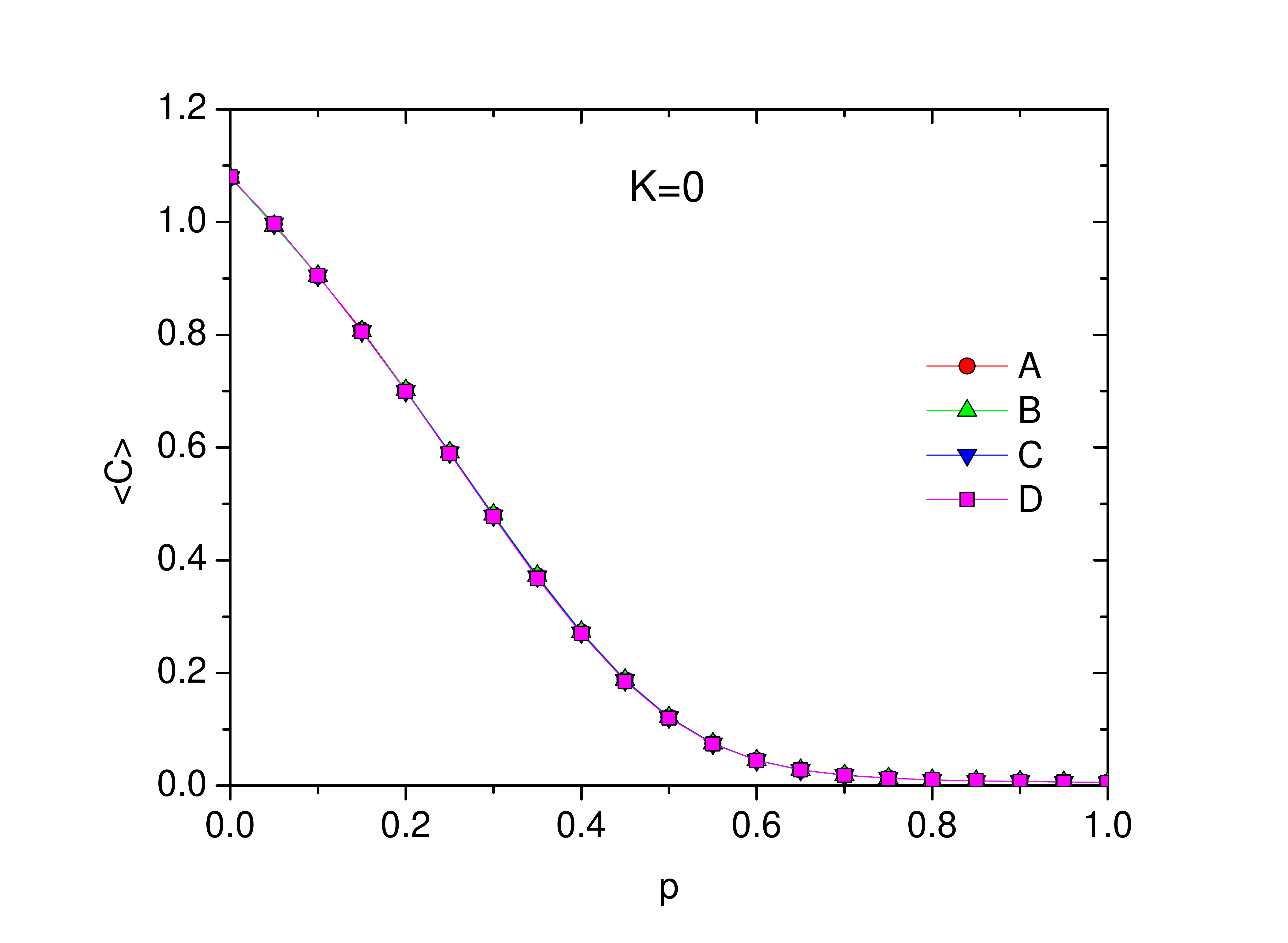}} \\
 	\subfloat {\includegraphics[width=0.48\textwidth]{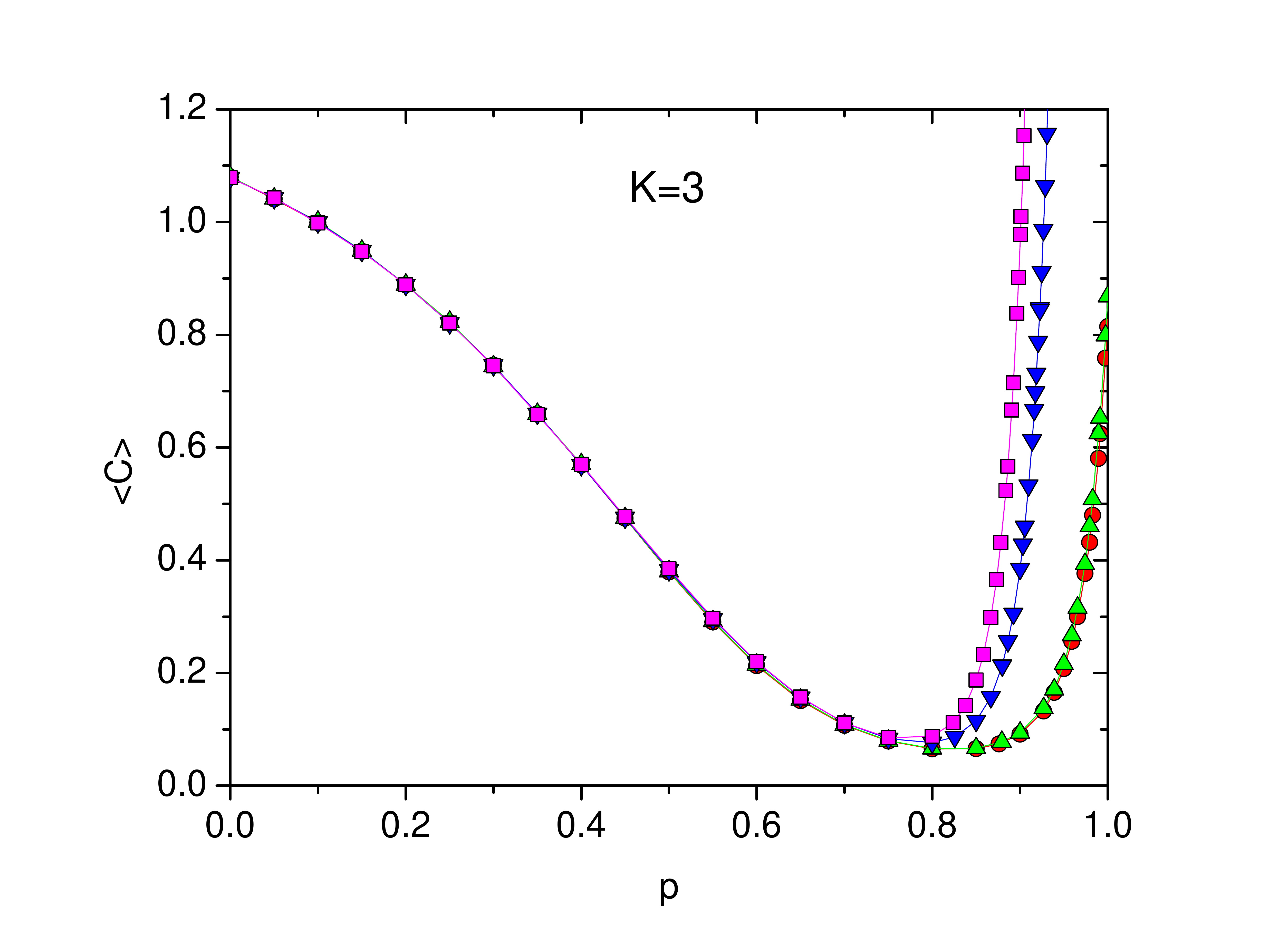}}\\
  	 \subfloat{\includegraphics[width=0.48\textwidth]{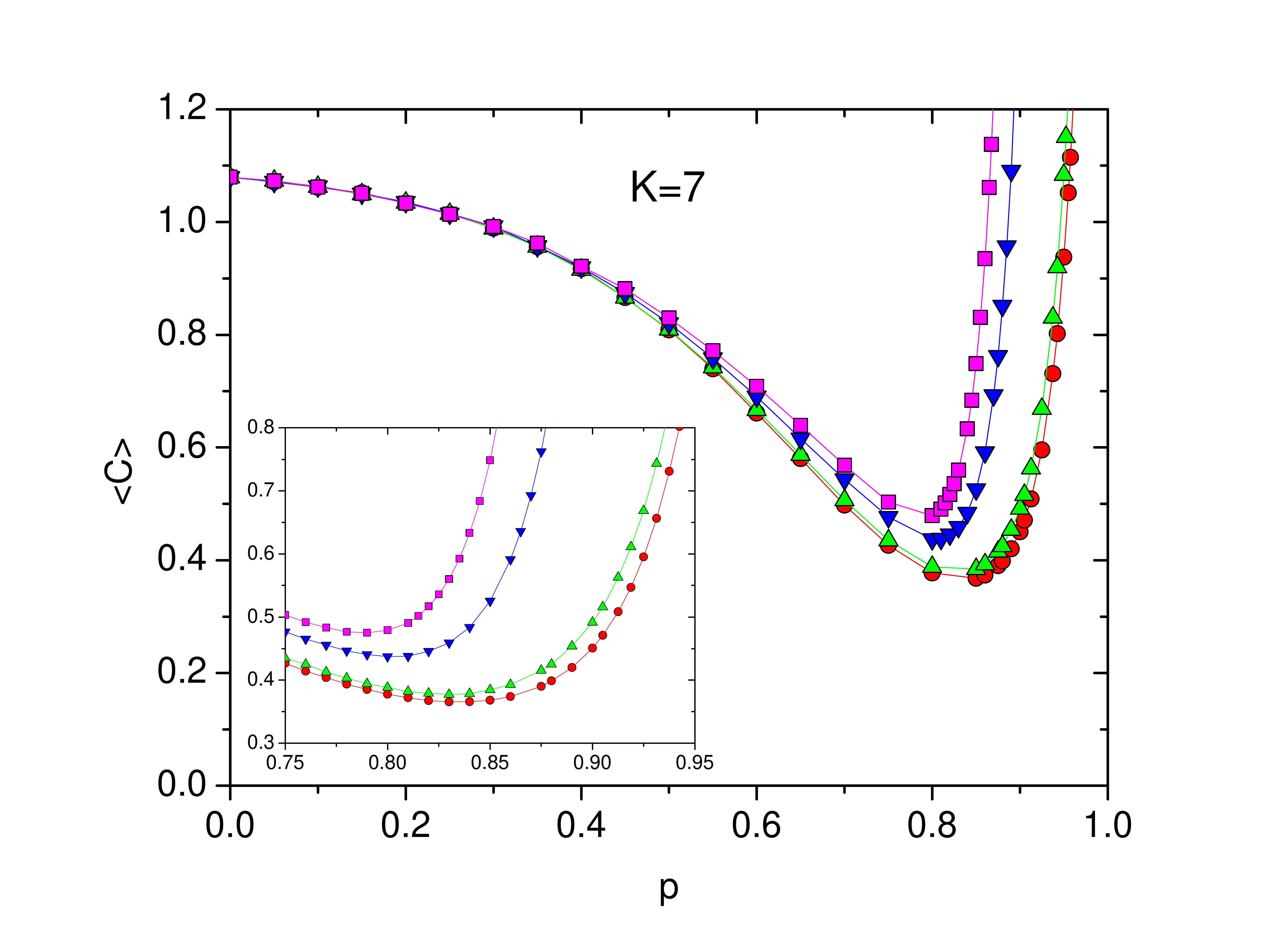}}
\caption{(Color online)  Average computational cost $\langle C \rangle$  as function  of the copy propensity $p$ for the four communication patterns shown in Fig.\ \ref{fig:1} according to the convention:  network A ($\CIRCLE$),  network B ($\blacktriangle$), network C ($\blacktriangledown$) and   network D  ($\blacksquare$). Each panel shows the results for a fixed value of the  parameter $K$ that measures the problem difficulty (see Table \ref{table:2}), from very easy ($K=0$) to difficult ($K=7$). The  lines are guides to the eye.
}
\label{fig:2}
\end{figure}

In Fig.  \ref{fig:2}, we show the dependence of the  average computational cost $\langle C \rangle $ on the copy propensity $p$ of the agents for increasing task difficulties as measured by the landscapes' ruggedness, $K=0$, $3$ and  $7$.  
For landscapes with no local maxima ($K=0$, upper panel of  Fig.  \ref{fig:2}), the performances of  the  four topologies  are practically indistinguishable in the scale of the figure, and the mean computational cost  decreases with increasing  $p$, i.e., the best performance is attained by always  copying the model string ($p=1$)  and allowing only its clones to explore the landscape through the elementary move. 

The  presence of  a moderate number of local maxima ($K=3$, middle panel of  Fig.  \ref{fig:2})  impacts the performance only for large values of the  copy propensity, and the effect is more pronounced for the more efficient networks C and D. The performances of networks A and B are very similar, except  in the region of  $p$ very close to 1, where network A slightly outperforms network B. The difference between the performances of these two networks becomes evident for difficult problems ($K=7$, lower panel of  Fig.  \ref{fig:2}) only,
as illustrated in the inset of that panel.

It seems that the determinant factor for the superior performance of a communication pattern is the variance of the betweenness  among the nodes (see Table \ref{table:2}). However, an alternative explanation may be the presence of modules in  network A and quasi-modular structures in networks B and C but not in network D, which exhibits the worst performance. In fact,  it has been argued that the modular organization, which is  characteristic of hierarchical networks, may facilitate the escape from the local maxima  \cite{Reia_16} (see \cite{Nagy_10} for experimental evidence  on the effect of a hierarchical social network structure  on  the  efficiency  of   collective  action). 

For the sake of concreteness, we  calculate the maximum modularity $Q$ of the networks shown in Fig.\ \ref{fig:1} in the case the nodes are assigned to two and three modules. We recall that the modularity of  a particular assignment of nodes into  modules (or communities) is defined as the fraction of the links that fall within the given module minus the expected fraction if the links were distributed at random \cite{Girvan_02,Newman_06}. Hence networks with high modularity have dense connections between the nodes within modules, but sparse connections between nodes in different modules. The maximum modularity is obtained by finding the assignment of nodes to modules that maximizes the modularity. For our networks, we  find that the maximum modularity occurs for the  partitioning of the nodes in three modules with the values $Q=0.581$ (network A), $Q = 0.565$ (network B), $Q=0.414$ (network C) and $Q=0.331$ (network D). This
supports the conjecture that the superior problem-solving performance of network A may be due to its high modularity. Actually, we have carried out an exhaustive search in the space of networks with $L=16$ and $k=3$ in order to determine the network with the highest maximum modularity value for partitioning of nodes in two and three modules and, as expected, the search produced network A.

In the case of rugged landscapes, the group performance correlates negatively with the efficiency of the networks, as measured by $\bar{l}$. This happens  because in this case, the model agents may broadcast misleading information, and so  it is advantageous to slow down the information transmission, so as to allow the agents more time to explore the solution space away from the 
neighborhoods of the local maxima.

We find that  $\langle C \rangle $ is quite insensitive to variations on the topology of the network  for small values of $p$. In particular, for $p=0$ one recovers the results of the independent search,  $\langle C \rangle \approx 1.08$ (see
\cite{Fontanari_15} for an analytical estimate of this value), regardless of the network topology and of the value of the parameter $K$. As the difficulty of the task increases, the  optimum copy propensity decreases towards zero, and the minimum of  $\langle C \rangle $ becomes shallower. In particular, for $K=N-1$ that minimum happens at $p=0$. 
Finally, we note that since finding the global maxima of NK landscapes  with $K>0$ is an NP-Complete problem \cite{Solow_00}, one should not expect that the imitative  search (or any other search strategy, for that matter)   would find those maxima for a large sample of landscapes much more rapidly than the independent search.

\begin{figure}[!ht]
  \begin{center}
\includegraphics[width=0.48\textwidth]{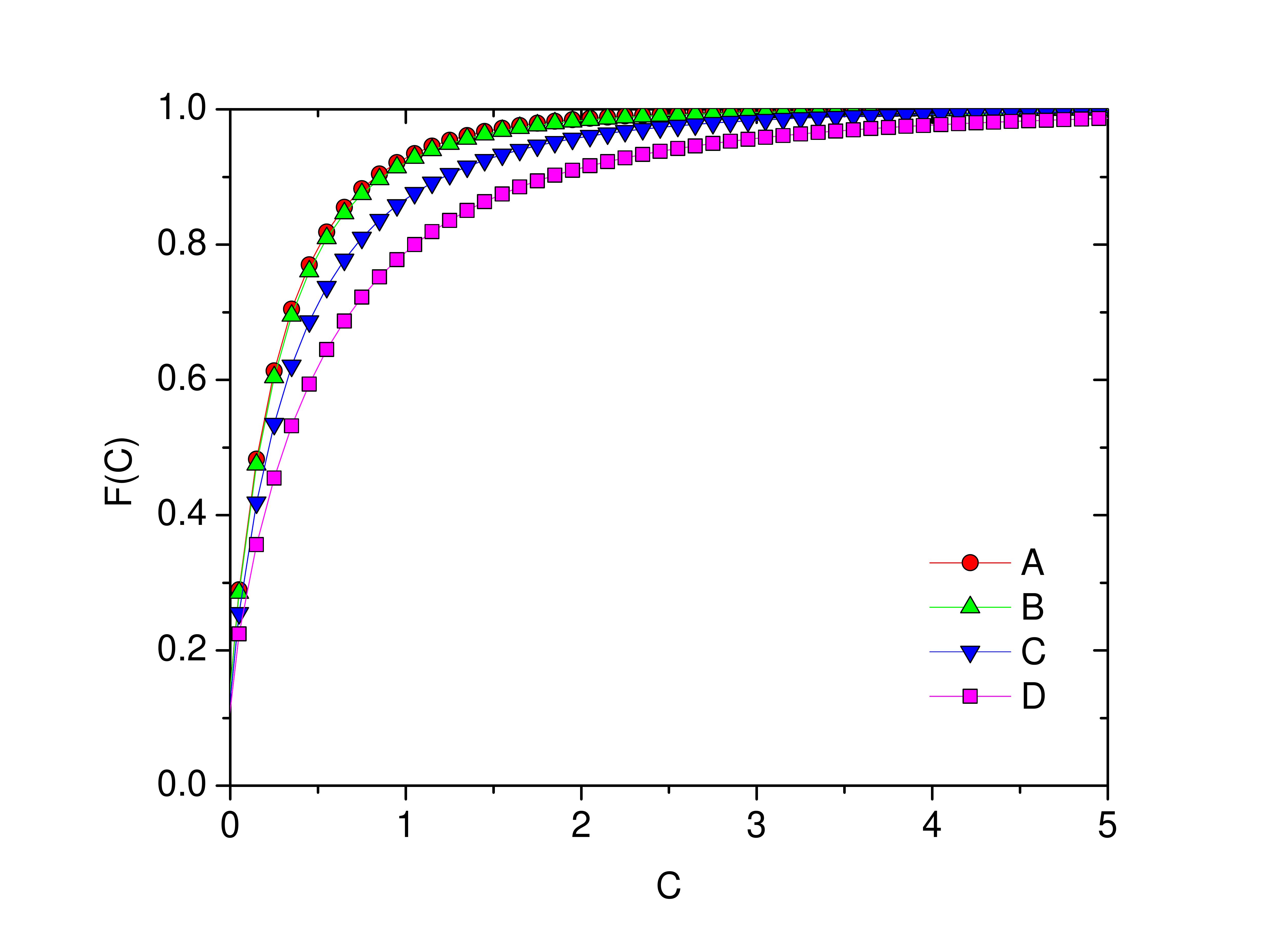}
  \end{center}
\caption{(Color online) Fraction of the searches that  found the solution for a fixed computational cost $C$    for network A ($\CIRCLE$), network B ($\blacktriangle$), network C ($\blacktriangledown$) and network D  ($\blacksquare$). The copy propensity is $p =0.85$ and the parameters of the landscape are
$N=16$ and $K=7$. The lines are guides to the eye.
 }
\label{fig:3}
\end{figure}

Figure \ref{fig:2} reveals the superior performance of networks whose nodes exhibit the largest variability of betweenness  in long runs, i.e., when there is no limit to the duration of the search. Now we  examine whether this finding holds also in the case where a maximum search time $t$ or computational cost $C$ is fixed a priori.  Figure  \ref{fig:3} shows the fraction of runs $F \left ( C  \right )$ that found the global maximum for a fixed value of $C$ in difficult tasks, i.e., landscapes with $N=16$ and $K=7$.  For easy tasks ($K=0$), we find that, similarly to the situation for long runs, the  fixed-cost performances of the four networks are practically indistinguishable (data not shown).
Hence, the conclusions for the long runs displayed in Fig.\ \ref{fig:2} are valid  in the case that the computational cost  of the search is
fixed a priori, as well.


\begin{figure}[!ht]
  \begin{center}
\includegraphics[width=0.48\textwidth]{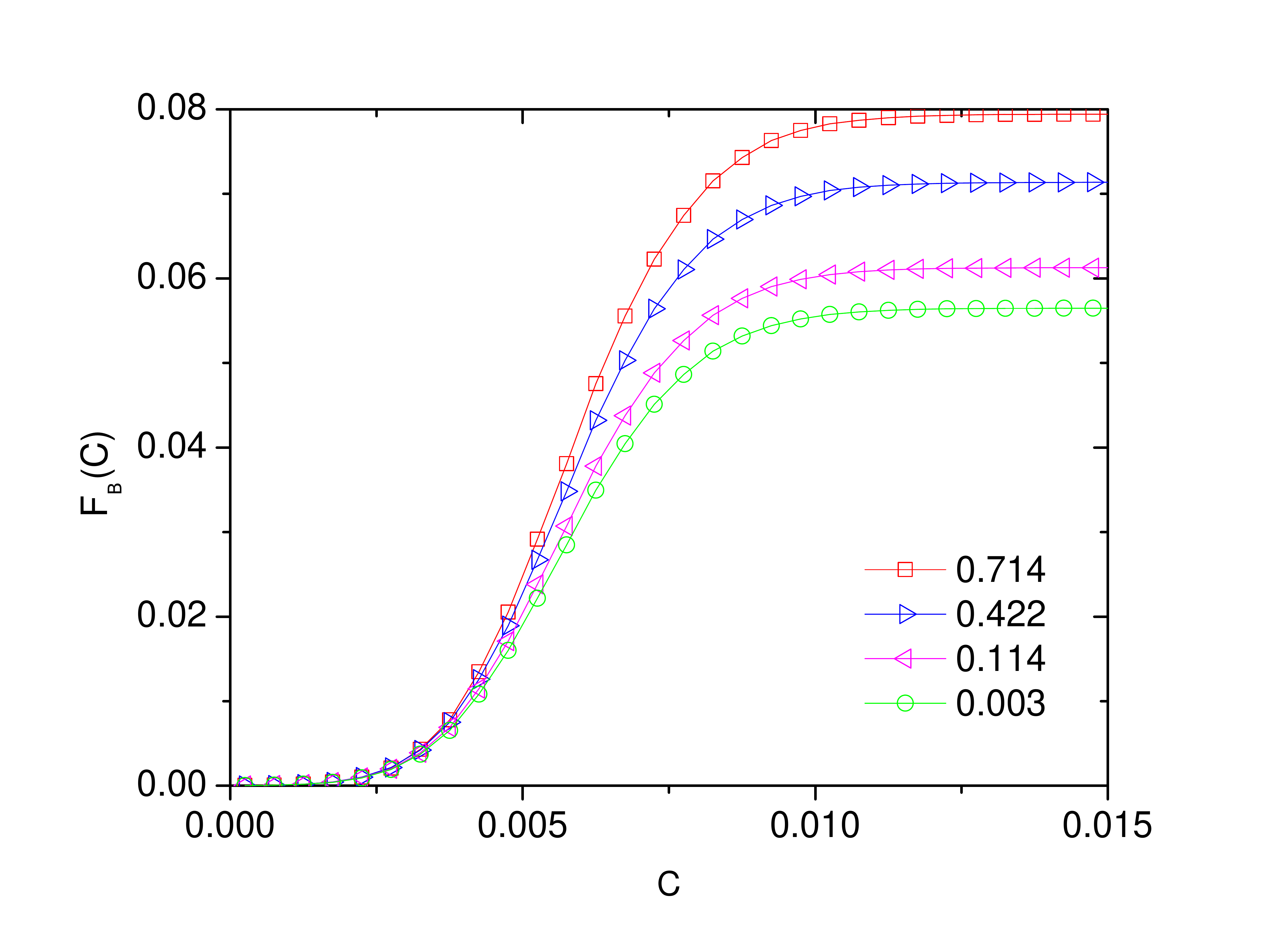}
  \end{center}
\caption{(Color online) Probability that a node  with betweenness $\mathcal{B}$ in network A  gets the answer first for runs with computational cost less than $C$. The convention is $\mathcal{B}= 0.00317 (\circ)$, $0.114 (\triangleleft)$,  $0.422 (\triangleright)$, and $0.714 (\Box)$ as indicated in the figure. The copy propensity is $p =1$ and the parameters of the smooth landscape are
$N=16$ and $K=0$. The lines are guides to the eye.
 }
\label{fig:4}
\end{figure}

\begin{figure}[!ht]
  \begin{center}
\includegraphics[width=0.48\textwidth]{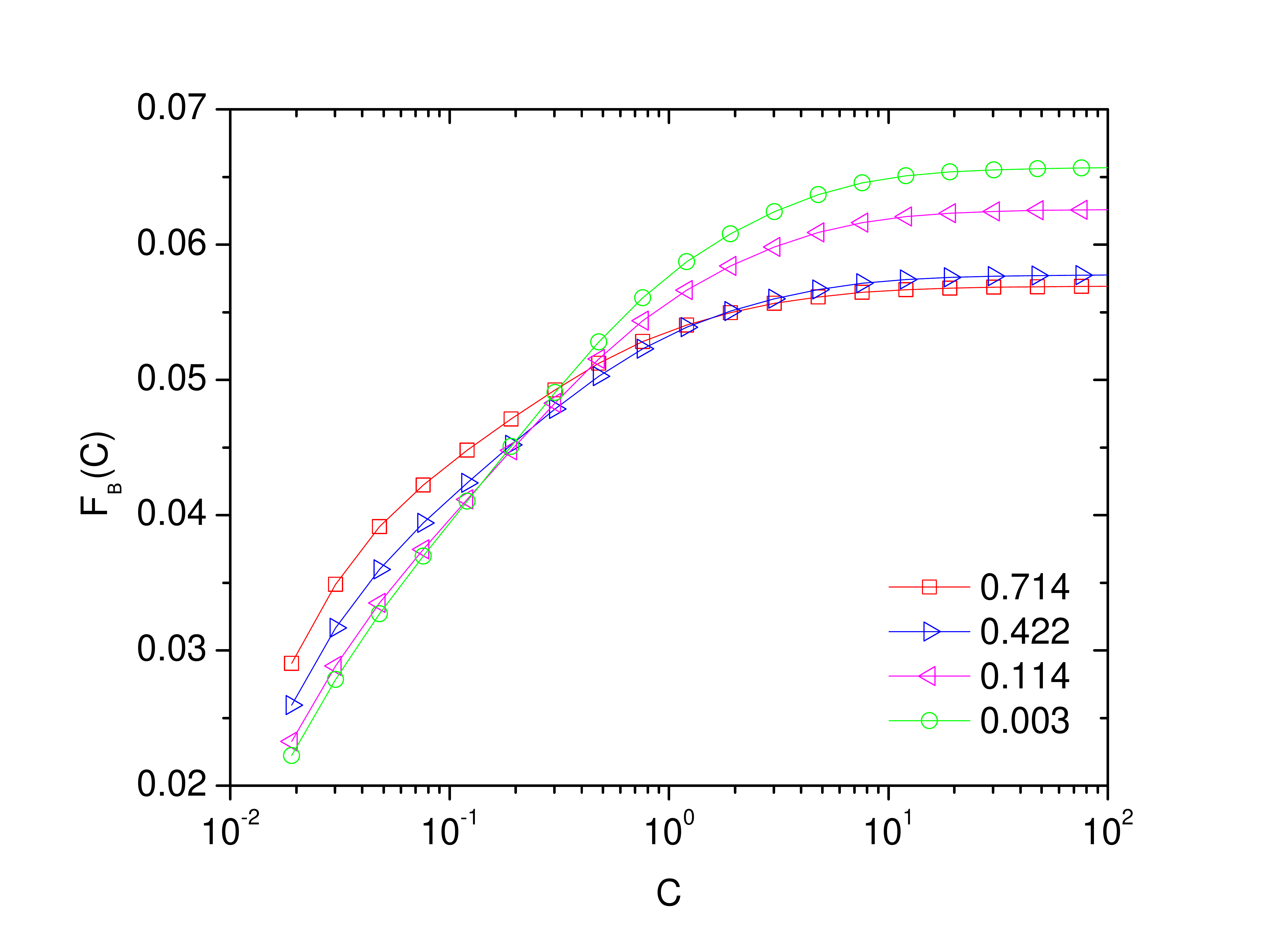}
  \end{center}
\caption{(Color online) Same as Fig.\ \ref{fig:4} but for rugged landscapes with parameters $N=16$ and $K=3$. The logarithmic scale in the x-axis highlights the change of the networks performance hierarchy  for long and short runs.
 }
\label{fig:5}
\end{figure}

For  network A, which exhibits nodes with 4 distinct betweenness values, namely, $\mathcal{B}= 0.00317$, $0.114$,  $0.422$, and $0.714$ with degeneracies $ D_{\mathcal{B}} = 6$, $6$, $3$ and  $1$, respectively,     
we can  measure the chance $F_{ \mathcal{B}} \left ( C  \right )$  that a specific node with a certain betweenness  finds the global maximum for a fixed computational cost $C$.
This  quantity is given by the fraction of runs with computational cost less than $C$ for which a particular node with betweenness $\mathcal{B}$ finds the solution. Since for $C \to \infty$, we can guarantee that all runs have halted, we have 
 $\sum_{\mathcal{B}} D_{\mathcal{B}} F_{ \mathcal{B}} \left ( C  \right ) = 1$  in this limit.
 The results for $K=0$ and $K=3$ are shown in Figs.\ \ref{fig:4} and \ref{fig:5}, respectively, with the copy propensity set to its maximum value, $p=1$.

For easy tasks ($K=0$), the central node of network A  is most likely to get the answer first, regardless of the allotted search time as shown in Fig.\ \ref{fig:4}.  More generally, the chance of a node  hitting the solution increases steadily with its betweenness. This conclusion holds for networks B and C as well, and for all values of the copy propensity $p$.  We recall that all nodes of  network D are identical regarding their betweenness values, so this network is unfit for this type of analysis.
  Surprisingly, the situation is reversed for more difficult tasks ($K=3$) as  shown in  Fig.\ \ref{fig:5}. Although it is still true that the central node is the most likely to find the solution for short runs (low computational cost), it is the least likely for long runs. In fact for long runs,  the lesser the centrality of a node, the greater its chance of  hitting the solution. We note, however,
the superior performance of the more central nodes for short runs seems to be a peculiarity of network A, since for networks B and C, the nodes with the lowest betweenness are the most likely to find the solution, regardless of the value of $C$ (data not shown).  It is interesting to mention a  related finding within the context of the spreading of epidemics in complex networks: under certain circumstances, the most efficient spreaders may not be the most central individuals in the network  \cite{Kitsak_10}.

\begin{figure}[!ht]
  \begin{center}
\includegraphics[width=0.48\textwidth]{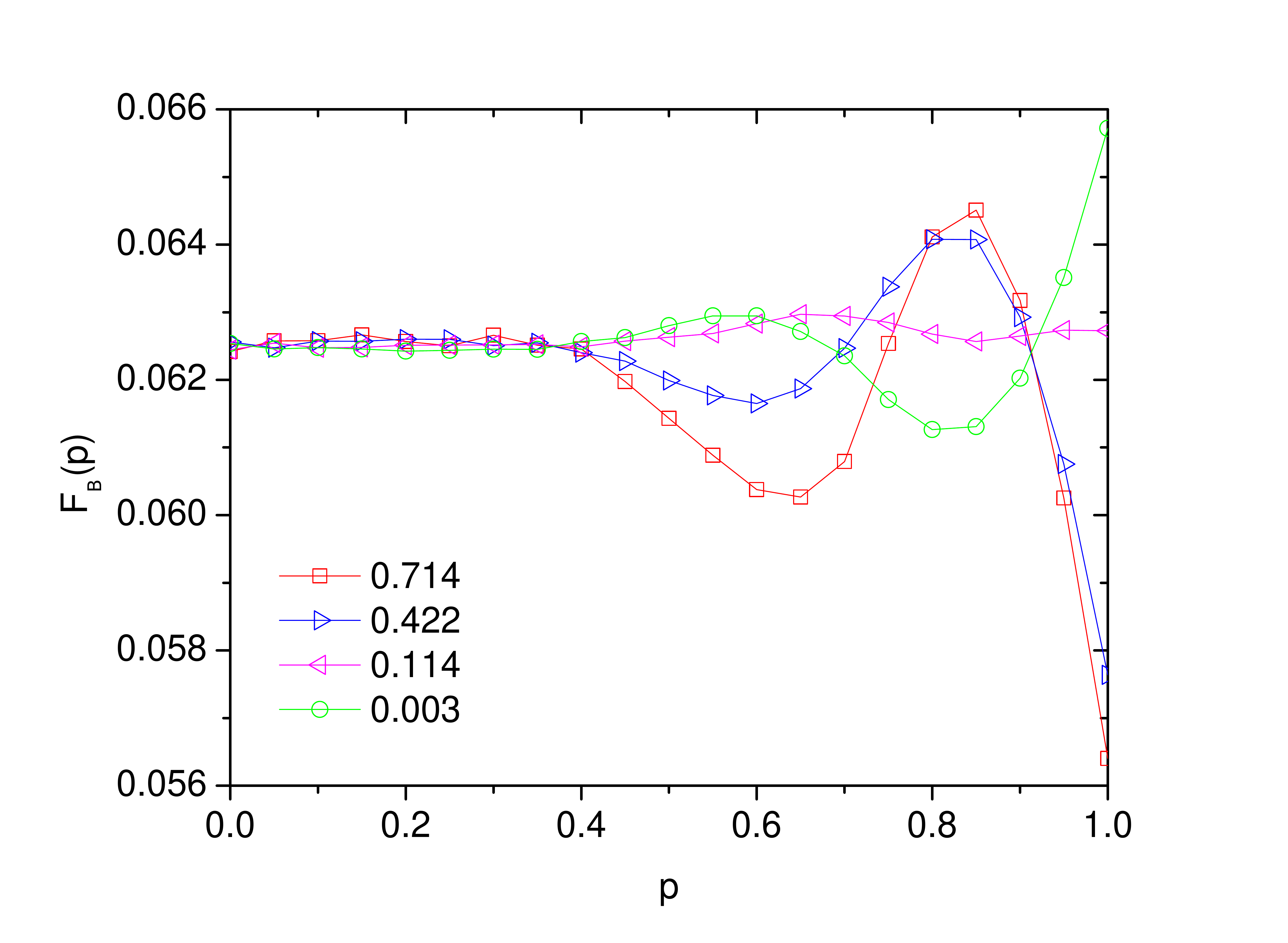}
  \end{center}
\caption{(Color online) Probability that a node  with betweenness $\mathcal{B}$ in network A  gets the answer first for time-unrestricted runs as function of the copy propensity. The parameters of the rugged landscapes are
$N=16$ and $K=3$. The symbols convention is the same as for Fig.\ \ref{fig:4} and the lines are guides to the eye. 
 }
\label{fig:6}
\end{figure}

\begin{figure}[!ht]
  \begin{center}
\includegraphics[width=0.48\textwidth]{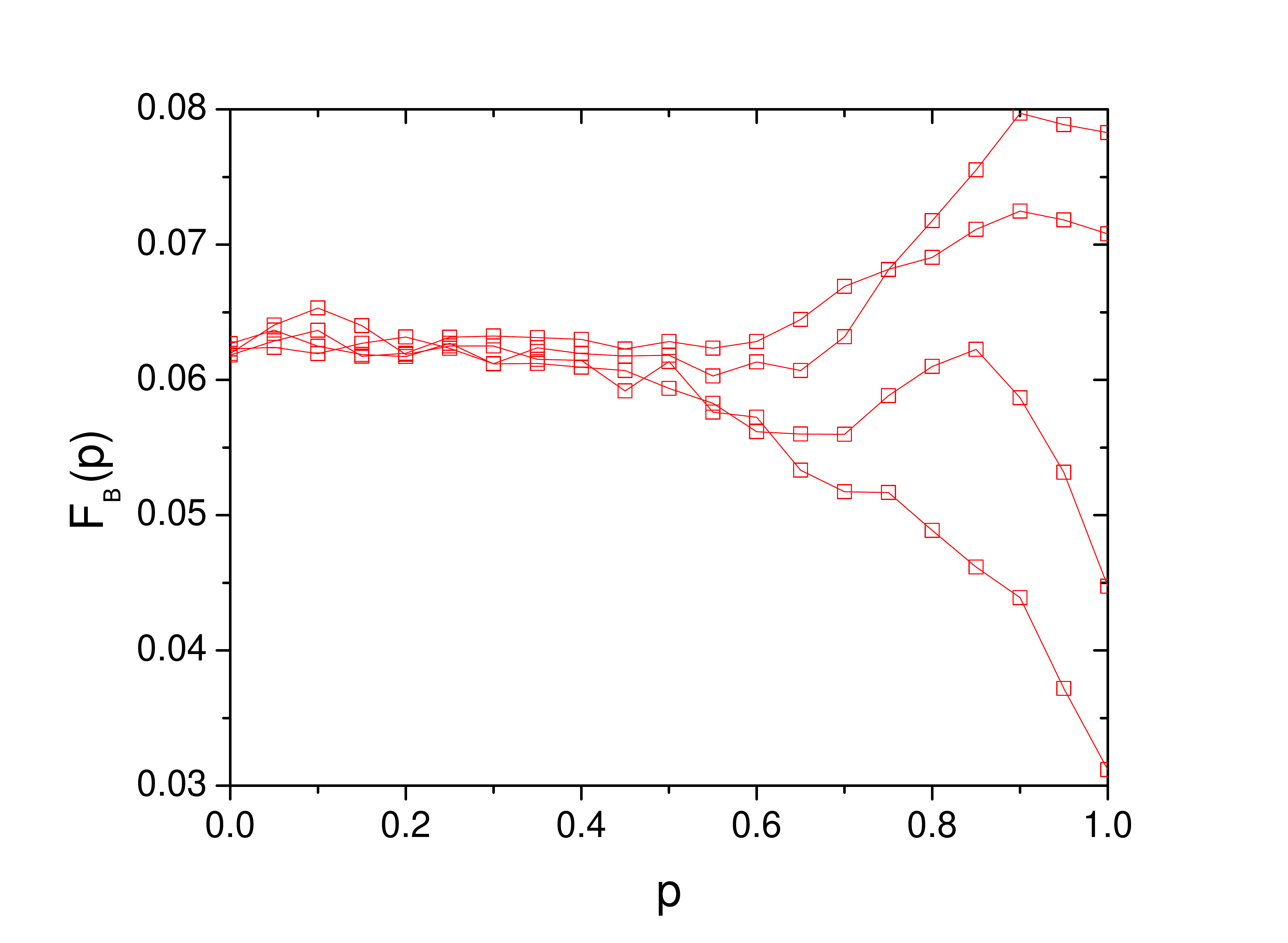}
  \end{center}
\caption{(Color online) Probability that the central node in network A  gets the answer first for time-unrestricted runs as function of the copy propensity  for four distinct landscape realizations. The parameters of the rugged landscapes are
$N=16$ and $K=3$ and the lines are guides to the eye. 
 }
\label{fig:7}
\end{figure}

Actually, for rugged landscapes the  performance of a particular node is way more complicated than for smooth landscapes,  because it depends on the copy propensity $p$ as shown in Fig.\ \ref{fig:6}. For instance, for $p$ close to the  value that minimizes the computational cost (see middle panel of Fig.\ \ref{fig:2}) the central nodes  perform better but for $p$ close to 1, in the region where the search is hampered by the local maxima, the peripheral nodes perform better, in agreement with Fig.\ \ref{fig:5}. We note that the highly non-monotonic behavior of the probability $F_\mathcal{B}$ for the central nodes is an artifact of   averaging over different landscapes realizations. In fact, Fig.\ \ref{fig:7} illustrates the strong  effect of the landscape realization on   $F_\mathcal{B}$ for the node  with the highest betweenness value (central node): for some landscapes, we find  that $F_\mathcal{B}$  decreases monotonously with increasing $p$ whereas for others,  this 
probability increases with $p$. The average over similarly discordant results using the sample of 100 landscapes  yields the  convoluted curves  exhibited in Fig.\ \ref{fig:6}.

Although we have found qualitatively similar results for NK landscapes with different ruggedness (i.e., different values of $K$),  there are two 
aspects that are worth mentioning. First, the differences in  the performances of the central  and peripheral nodes  for short runs become less noticeable with increasing $K$, and second, those differences become more  prominent  for  long runs. For instance, whereas the chances of hitting the global maximum are practically indistinguishable for nodes with betweenness  $\mathcal{B}= 0.714$ and 
$\mathcal{B}= 0.422$ for $K=3$ (see Figs. \ref{fig:5} and \ref{fig:6}), we found that the more central node significantly outperforms the less central nodes  for $K>3$. In summary, increase of $K$ decreases the effect of the centrality of the nodes for short runs, but increases it for long runs.

Since even  our simple agent-based model yields plenty of discordant results regarding the performance of individual nodes on difficult tasks, the profusion of conflicting conclusions drawn from the studies with human subjects is no wonder: neither the cooperation strategies used by the subjects, nor the difficulty of the problems are controlled variables in those experiments.

 \section{Discussion}\label{sec:disc}

The claim that  restrictions on the communication channels available to a group affects its problem-solving efficiency is hardly controversial. 
However, the issue whether there is  a communication pattern which  gives significantly better performance  than others in solving  specific or general  tasks has produced conflicting findings \cite{Bavelas_50,Leavitt_51,Heise_51,Guetzkow_55,Shaw_54,Mulder_60,Mason_12,Lazer_07,Herrmann_14}. The main reason seems to be the strong dependence of the group performance on several aspects of the group organization, as well as  on the complexity of the tasks  and on the cooperation strategies used by the subjects. 
Hence a thorough
analysis of the vast parameter space  of the  group/task composite   is  necessary to elucidate that issue. Such a  comprehensive study is feasible only through computational experiments \cite{Lazer_07}.
  
Here we attempt to clarify the role of the betweenness centrality on the performance of a group as well as on the performance of its individual members. To achieve that, we compare networks with the same number of nodes ($L=16$), which are identical with respect to their degrees ($k=3$ for all nodes), as illustrated in Fig.\ \ref{fig:1}. The networks were generated so as to exhibit special properties regarding their local and global betweenness metrics \cite{Mason_12}. 
Actually, the reason we used such small networks, in addition to the need to compare our findings with those of Ref.\ \cite{Mason_12} as will be done below, is that  it is unfeasible to find networks with those special properties through the exhaustive search in the space of networks for a number of nodes larger than  $L=16$.
The task posed to the agents is to find the unique global maximum of a NK fitness landscape with  the parameter $N=16$ fixed but with variable $K$. The difficulty of the task increases with $K$ due to the proliferation of local maxima.
The choice $N=L=16$  ensures  that the group size is fixed close to its optimal  value  for easy tasks \cite{Fontanari_15}, and that the size of the state space $2^N$ is much larger than $L$, which  makes the search for the single global maximum a challenging task.

We find that for simple tasks, there  is no significant difference in  the performances of the different network topologies. However, the performance of the group members, which is  measured by the chance they get the answer first, is strongly correlated to the centrality of the node  in the network: the more central a node is, the more likely it is to find the solution of the task. These findings are in agreement with the
experimental results  \cite{Leavitt_51,Shaw_54}.  We stress, however, that the unresponsiveness of the easy-task performance to changes on the communication patterns is because the number of nodes, as well as the density of links are  the same  for the four networks of Fig.\ \ref{fig:1}.  Variation of these parameters has a
large effect  on the group performance for easy tasks. For instance, for those tasks there is an optimal group size that minimizes the cost of the search,  and the optimal performance is achieved by  fully connected networks  \cite{Francisco_16}, in agreement  with the experimental finding that groups with more communication channels perform better for simple tasks \cite{Heise_51}.

For complex tasks, the best performing  communication pattern is the pattern that maximizes the variance of the betweenness among the nodes
(network A in  Fig.\ \ref{fig:1}). This network also exhibits  the node with the largest possible betweenness  allowed  by the constraints 
$L=16$ and $k=3$  for all nodes. The group performance degrades as that variance decreases towards zero, so that the worst performing pattern is  that in which all nodes have the same betweenness (network D in  Fig.\ \ref{fig:1}). Interestingly, among the topologies displayed in Fig.\ \ref{fig:1}, network D is the topology with the shortest average path length, and so it is the most efficient regarding the transmission of information through the network, whereas network A is above  network B only in this rank of efficiency (see Table \ref{table:2}).
 Our finding that inefficient networks perform better  is justified by noting that speeding up the transmission of information through the network makes sense only if one can guarantee its faithfulness and usefulness, otherwise it may be wiser to slow it down  and give more time for the agents to explore different regions of the state space
 \cite{Lazer_07,Mason_08,Fang_10}.   We note, however, that even in the case the fitness values provide faithful information on the location of the global optimum, viz. for landscapes with $K=0$, the so-called efficient networks do not perform better than the inefficient ones.

 In addition, the attempt to maximize the betweenness of all nodes results in the network with the largest average betweenness   (network B in  Fig.\ \ref{fig:1}), which exhibits the second best performance. Thus it seems that the key factor to improve group performance is not the maximization of the number of nodes that have large betweenness,  but rather the assignment of a large variety of betweenness values to  the nodes. Hence, within the perspective of the  betweenness centrality metric,  diversity is crucial to boost  group performance.

Regarding the performance of an individual node of the communication network, measured by the probability that it hits the global maximum first,  we find a neat positive correlation between the centrality of a node and  its performance  on easy tasks, regardless of the topology or of the particularities of the agents, such as their copy propensities $p$. For complex tasks, however, there is no such a general verdict as the  performances of the nodes are strongly influenced by the agents dispositions to cooperate  and by the specific realizations of the rugged landscapes. Averaging over the landscape realizations yields a complicated dependence  on the parameter $p$. For instance, 
the central nodes perform better when the network performance is optimal, whereas the  peripheral nodes win when the network performance
is most heavily harmed by the local maximum traps. This sensitivity on the details of the model and the consequent impossibility of drawing general (i.e, task and subject independent)  conclusions is reminiscent   of the conflicting outcomes  that characterize the  experimental literature dealing with the effects of the communication patterns in task-oriented  groups \cite{Bavelas_50,Leavitt_51,Heise_51,Guetzkow_55,Shaw_54,Mulder_60,Mason_12,Mason_08,Fang_10}.

A word is in order about the interesting online experiments conducted using Amazon's Mechanical Turk,   in which  human subjects (players)  select  points for oil-drilling on a map \cite{Mason_12}.  The good and the bad oil wells are the maxima and the minima of a rugged landscape and their locations are unknown to the subjects, who, however, are  able to see the coordinates of the selected points as well as the  earnings of their network neighbors. The goal of each player is to maximize its own earnings  by picking the more productive well.  In  contrast with our findings, the best performing networks in the web-based experiments are the so-called efficient networks, characterized by short average path lengths, such as networks C and D in Fig.\ \ref{fig:1}. A possible  reason for this discrepancy is the distinct  performance measure used in those experiments, namely, the average earnings of the group rather than the time to find the optimum oil well. In addition, since the game does not stop when that optimum is eventually  found by a player,  truly useful information about the coordinates  of the optimum becomes available to the other players, which may then explain the superiority  of the topologies with short average path lengths.

Finally, we note  that the study of distributed cooperative prob\-lem-solving systems diverges from the game theoretical literature on cooperation that followed Robert Axelrod's 1984  seminal book The Evolution of Cooperation  \cite{Axelrod_84}. In fact, in the context of 
cooperative processes,  there is no conflict of interests between the agents, and  the opposite of cooperation is independent work, rather than defection. In addition, in the game theoretical framework it is usually assumed that  mutual cooperation is the most rewarding strategy for the group in the long run, whereas here we  argue that too much  cooperation, which results from a high value of the copy propensity, may lead to disastrous results, akin to the so-called groupthink phenomenon that happens when everyone in a group starts thinking alike \cite{Janis_82}.

Most studies of the influence of  communication patterns on the operation of groups have considered externally imposed patterns that dictate the distribution  of  the  communication channels among the group members (see \cite{Guetzkow_55} for an exception), thus precluding the emergence of a self-organized network. This  connectivity-driven network approach is fitting to describe situations where the connections among nodes are persistent features, such  as in the Internet, but it may not be so suitable to model the more ephemeral work relationships, which  may change on a very short time scale.  A promising avenue for further theoretical investigation is to consider dynamic unidirectional links,
as in activity driven models of varying networks \cite{Perra_12},
 that can be created or destroyed depending on whether a previous copy process resulted in an increase or decrease of the  copier fitness. 
 The topology of the resulting time varying self-organized network may shed light on the issue of the emergence of leadership in a task-force
 \cite{Lindquist_09}.

\acknowledgments
The research of JFF was  supported in part by grant
15/21689-2, S\~ao Paulo Research Foundation
(FAPESP) and by grant 303979/2013-5, Conselho Nacional de Desenvolvimento 
Cient\'{\i}\-fi\-co e Tecnol\'ogico (CNPq).  SMR  was supported by grant  	15/17277-0, S\~ao Paulo Research Foundation (FAPESP).

\end{document}